**Electrohydrodynamic self-boosted propeller for in-atmosphere propulsion**

Adrian Ieta[1] and Marius Chirita[2]

Attempts to use Electrohydrodynamic (EHD) flow for propulsion have been made since the last century[1]. Limited success has been registered particularly due to the inhomogeneous generation of the thrust, and also the very light weight and frail nature of the devices versus the power supply weight needed hover the craft[2]. However, EHD propulsion can offer a greater thrust to power ratio than any of the current propulsion technologies[3]. Rotary EHD devices have been employed as demo units for a long time[4,5], but it was unknown if they could produce enough thrust and vertical lift to lead to device liftoff.  We designed EHD propellers which spin and eventually lift off and fly independently for a short while. The propeller is balanced on and powered through a high voltage pin/shaft while an intense electric field is created by the presence of a surrounding ground electrode[6]. Multiple propeller electrode designs and ground counter electrodes are able to support propeller rotation and liftoff. Propellers up to 27.8 g and 25.5 cm in diameter were studied with a liftoff voltage range of negative 9.5 kV to 60 kV and rotational speeds up to 80 rot/s. It appears that these are the first EHD devices to liftoff and fly on their own without carrying a power supply[7]. This validates a rotary-based system as a new principle for propulsion with potentially improved stability and control characteristics versus the current EHD devices. Design optimization and scaling seem possible and EHD drone design feasible. Other applications may include EHD motors, fans, or sensors.

**Introduction**

EHD flow is generated as a result of local ionization of air near sharp electrode edges when high voltage above corona onset is applied. The ions accelerate in the electric field and transfer momentum to neutral atoms in a charged particle–neutral coupling process[8] which is only partly understood. The generated air flow is also known as corona, ionic, or electric wind. Ion speed is proportional to the local electric field and it is about two orders of magnitude higher than the induced bulk air speed[8].  An initial evaluation of the ionic wind for propulsion was conducted by Brown[1] and later on was substantiated by Christenson and Moller[9]. Most EHD devices use asymmetrical capacitor wire-to-plane configurations and the resulted downward EHD thrust. Renewed interest in the topic has been shown by NASA[2,10,11] and other groups[3,12-16]. Recently, good progress has been made in the miniaturization of EHD propulsion devices and in the decrease of the operating voltage[15,16]. Rotational motion has rarely been employed in the study of ionic wind, although a two stage rotating pinwheel based on the phenomenon was first developed by Wilson in 1750[4]. However, the first device with self-induced rotary motion by corona wind was developed in 1760[5]. An EHD rotary device was employed by Canning *et al.*[10] for finding the EHD thrust generated by asymmetric capacitors. Lately, a NASA research grant report presents the determination of the EHD forces for a hobbyist triangular electrostatic lifter substantiating the rotational motion analysis and air drag[17,18]. Our previous work on EHD flows in rotary setups[19-21] made use of pin emitters as actuators for EHD spinners. In the present work we hypothesized that a corona wind-activated rotary propeller device may be able to spin fast enough in air at atmospheric pressure to liftoff at least its own weight. We focused on testing this assumption using already flight optimized propellers and various setups.

[1] Electrical and Computer Science Department, SUNY Oswego, 7060 State Route 104 Oswego, NY 13126-3599, USA.

[2] Condensed Matter Department, National Institute for Research and Development in Electrochemistry and Condensed Matter, Plautius Andronescu Str. 1, RO-300224, Timisoara, Romania.
Correspondence and requests for materials should be addressed to A.I. (email:ieta@oswego.edu).

**Theoretical background.** When ionic wind is produced, EHD thrust is generated from the action of the electric field on the space charge and as a result of momentum conservation for the air-electrode system. If the electric field $E$ and spatial charge density $\rho$ distributions in the volume $v$ are known, the thrust $T_{EHD}$ can be calculated as

$$T_{EHD} = \int \rho E dv \qquad (1)$$

In static setups, the thrust is proportional to corona current[2,22,23]. Wilson *et al.*[2] found experimentally that a more realistic relation modeling the thrust is

$$T_{EHD} = I \frac{kd^n}{\mu} \qquad (2)$$

where "$I$" is the corona current, "$d$" is the distance between the corona wire/point and the ground electrode, "$k$" is a proportionality constant, $\mu$ is the average mobility of ions, and n is a constant coefficient with n=1 for small currents or less than one otherwise. Corona current is generally accepted to follow a quadratic variation with the voltage above the corona onset [24, p.261]

$$I = KV(V - V_o) \qquad (3)$$

where $V_o$ stands for corona onset voltage and K is a proportionality constant. Therefore, the thrust should follow a quadratic variation with the applied voltage for constant "d"

$$T_{EHD} = BV(V - V_o) \qquad (4)$$

where B is the new resulting constant. If EHD thrust is applied on the blades of a propeller, it generates torque and the resulting rotational motion can be characterized by

$$J_p\ddot{\theta} + C\dot{\theta}^N = \tau_{EHD} - \tau_{F_f} \quad (5)$$

where "$J_p$" is the moment of inertia of the propeller, C is a constant, $\dot{\theta}$ is the angular speed, $\ddot{\theta}$ is the angular acceleration, $\tau_{EHD}$ is the torque associated to the EHD generated thrust, $C\dot{\theta}^N$ is the torque associated to the drag forces with N=1 for laminar flow and N=2 for turbulent flow, and $\tau_{F_f}$ is the torque associated in our setups to the frictional forces between the axial support needle and the propeller.

The thrust efficiency is defined as

$$\eta = \frac{T_{EHD}}{IV} \quad (6).$$

In conjunction to relation (2), it shows the tradeoff between efficiency of thrust generation and the applied voltage.

**Results**

**EHD propeller.** Rotational EHD devices were produced by redesigning commercially available mini-drone propellers. Any conductive sharp point or edge can generate EHD flow when in an intense electric field. An electrode was designed on the propeller blades by attaching thin conductive material to at least one blade edge, starting from its axial part (Fig1b). The conductive material used included one or more of the following: thin metal tape, metal pins, thin wire, and conductive ink. A dielectric layer was used to partially cover the conductive material, occasionally leaving the edge exposed to air at the trailing edge of the blade. The dielectric layer partly blocks the EHD flow generation at the covered metal edge while allowing it to form freely at the uncovered side. Both pins and metal tape produce corona wind and therefore thrust and EHD torque $\tau_{EHD}$. An axial metal pin is used to inject high voltage (HV) in the blades through the conductive material running along the blades. As the propeller moves and rotational speed increases, the vertical thrust cancels the normal force on the balancing pin or metal bead, eventually reducing the rotational frictional torque to zero if liftoff is achieved. A surrounding ground electrode helps create an intense electric field. In some designs, metal pins (Fig. 1a) or pointed protrusions extend from the trailing edge of one or more blades. By applying high-voltage above corona onset to the supporting metal pin or rotor shaft electrode, ionic wind is produced causing the rotational device to spin. Using the EHD generated flows, we aimed at and obtained liftoff for many propellers. Propeller liftoff was achieved using different electrode designs: metal tape covered partly by insulating electrical tape with or without pins on one or more blades, conductive wire covered partly by insulating tape along the one or more of the blades, and conductive ink along the trailing edges of blades with or without pins. The addition of one or sometimes two pins per blade usually lowered the liftoff voltage by a few kilovolts.

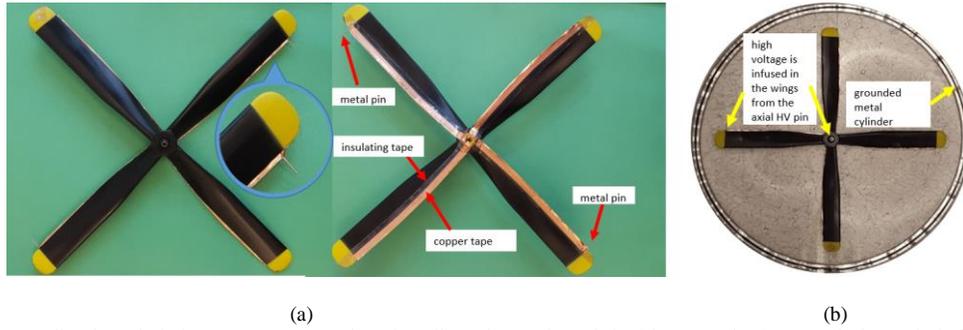

(a) (b)

Fig.1 (a) Sample propeller electrode design; copper tape runs along the trailing edges, and metal pins (about 1 cm long) are attached towards the blade tips; the copper tape inner edges are covered by transparent insulating tape partly blocking the ion emission in that direction (b) sample propeller setup (top view): the propeller was usually balanced axially on a high voltage metal pin and placed in the central part and along the cylinder axis where the most intense electric field is produced.

| Propeller # | Mass [g] | Number of blades | Diameter [cm] |
|---|---|---|---|
| 1 | 0.198 | 2 | 3.5 |
| 2 | 0.462 | 3 | 4.6 |
| 3 | 0.483 | 2 | 5.5 |
| 4 | 0.582 | 3 | 4 |
| 5 | 2.233 | 2 | 12.6 |
| 6 | 2.708 | 2 | 24 |
| 7 | 8.026 | 6 | 12.5 |
| 8 | 9.81 | 2 | 25.5 |
| 9 | 11.581 | 2 | 23 |
| 10 | 27.791 | 4 | 25 |

Table 1. Basic characteristics of redesigned propellers types that reached liftoff. Propeller # is used for identification of the propeller type throughout this article. The mass of the propellers in the actual experiments may vary slightly according to the particular design used.

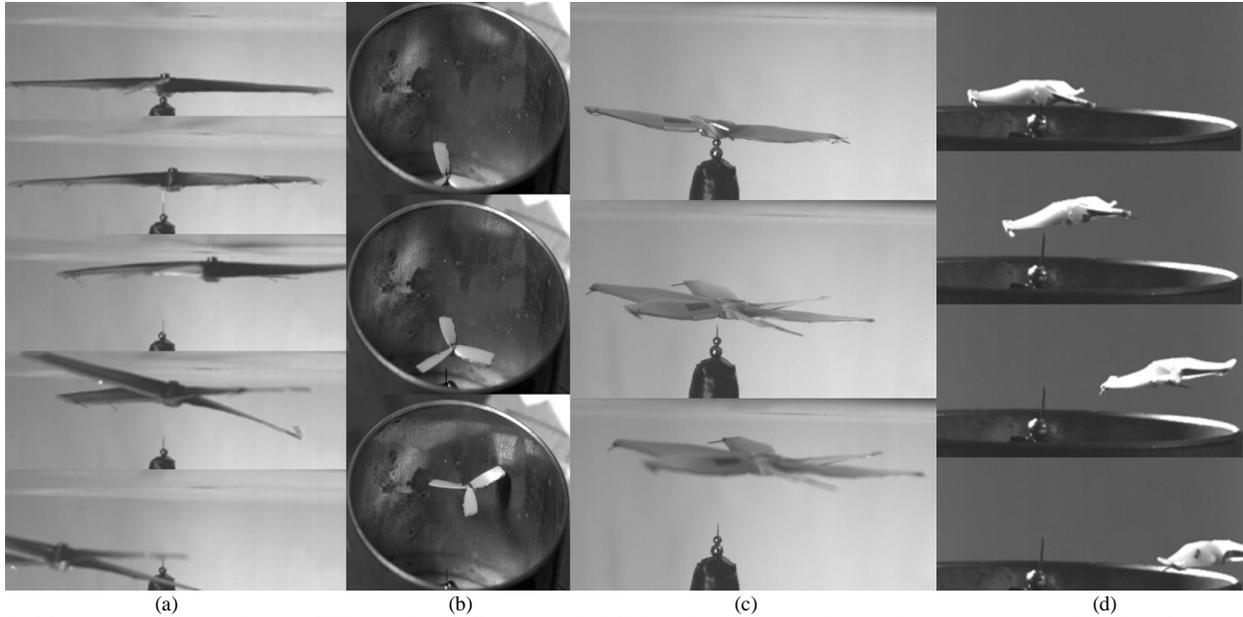

(a) (b) (c) (d)

Fig.2 Sample sequences (top to bottom) of EHD propeller liftoff and unrestrained flight. (a) Propeller (10) designed with copper tape on the trailing edges of the blades and two pins per blade; flight induced at -52 kV, 4.7 cm below a centered 61 cm diameter ground disk electrode; the four-blade propeller lifts off and bumps into the disk above it, suggested also by the impact wave visible on the blades on the third imagine in the sequence (10,000 fps). (b) propeller (2) designed with conductive ink on the trailing edges of the blades; flight induced at -28.9 kV in a 10.5 cm diameter, 11 cm height copper cylinder (6,000 fps). (c) Propeller (7) designed with copper tape on the blades and one pins per blade; flight induced at -60 kV, 4.7 cm below a centered 61 cm diameter ground disk electrode (10,000 fps). (d) Propeller (4) designed with copper tape and one pin on the trailing edges of the blades; flight induced at -32 kV in a 10.5 cm diameter, 11.2 cm height copper cylinder (6,000 fps).

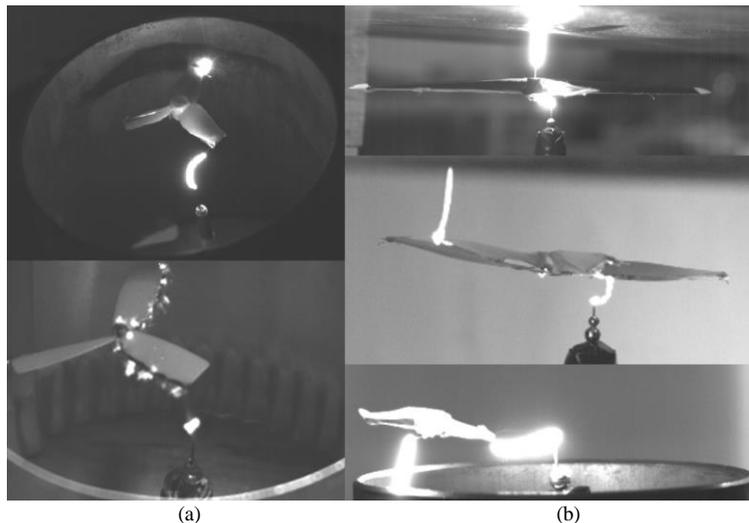

(a) (b)

Fig 3. Sample breakdown discharges from the propeller during unrestrained flight. (a) Top: propeller (4) designed with copper tape and one pin per blade; discharge at -23.4 kV in a 10.5 cm diameter, 11 cm height copper cylinder (30,000 fps). Bottom: propeller (2) designed with conductive ink on the trailing edges of blades - at -27 kV in a 10.2 cm diameter, 9 cm height aluminum cylinder. Multiple discharges along the conductive edges are noticeable due to the burning ink particles detached from blade edges (10,000 fps). (b) Top: propeller (10) at 50 kV and initially 4.7 cm below a centered 61 cm diameter ground disk electrode. Middle: propeller (7) at 60 kV and initially 4.7 cm below a centered 61 cm diameter ground disk electrode (10000 fps). Bottom: propeller (4) designed with copper tape and one pin on the trailing edges of the blades; flight induced at -32 kV in a 10.5 cm diameter, 11.2 cm height copper cylinder (6,000 fps).

**Propeller Flight.** An EHD propeller was used in conjunction with a grounded counter electrode which consisted of a metal or semitransparent aluminum screen cylinder, parallel metal plates, and a metal disk. Liftoff was obtained for different propellers inside, below, and above the grounded cylinder, between grounded parallel metal plates; below a horizontal metal disk, and near external wall of a metal cylinder. Their basic characteristics are given in

Table 1. Both positive and negative polarities were tested, but liftoff has only been obtained in the negative polarity. Measurements for the propeller (4) – Table 1- showed that corona current in negative polarity was consistently larger than in positive polarity given the same voltage magnitude applied. Also, breakdown discharges happened at much smaller voltages in the case of positive polarity than in the negative one, limiting the amount of EHD thrust to be generated. Sample propeller liftoff time sequences are shown in Fig.2. Propeller (10) in Fig. 2a lifts off and bumps into the above counter-electrode disk. Propeller (8) in Fig. 2b, flies higher from the central area of the cylinder where the electric field is more intense. The 6-blade propeller (7) has more of a stable sideways flight (Fig. 2c) with many partial discharges between the blades and the central pin. As propeller (2) lifts off the HV pin (Fig. 2d), the field intensity decreases and the EHD force and torque diminish as well. The flight is still induced though it is generally more difficult to obtain. Breakdown discharges typically follow during the unrestrained propeller flight (Fig.3) as it approaches the counter-electrode and the breakdown electric field intensity is reached.

**Limiting factors and performance**. The voltage at liftoff depends on various parameters. The minimum distance between the propeller electrode and the counter-electrode sets the electric field intensity and thus the thrust – equation (1). This accordingly sets the upper limit of the applied voltage to the air breakdown voltage for that distance (no significant thrust is generated when air breakdown happens). With a 27.791g mass and 12.5 cm radius, propeller (10) was the heaviest tested and was able to liftoff at negative 42.9 kV (Fig.2a). With a 0.198 g mass and 1.75 cm radius, propeller (1) is the lightest one tested. The lowest voltage it was able to detach from the HV pin at (among all tests performed) was negative 9.5 kV. The voltage values needed for liftoff are not absolute since the intensity of the electric field ultimately determines the EHD forces and torque produced. For instance, propeller (4) detached from the HV pin and flew at negative 54 kV in a metal cylinder 29 cm in diameter and 60 cm in height while a liftoff was also obtained at negative 16.9 kV in a cylinder 7.5 cm in diameter and 5.4 cm in height.

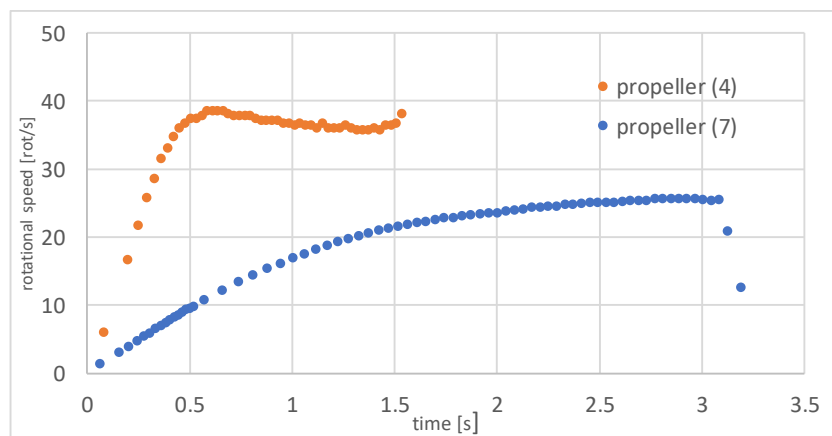

Fig 4. Sample liftoff regimes: (a) liftoff of propeller (4) designed with copper tape and one pin per blade; the applied voltage was -25.8 kV in a 10.5 cm diameter, 11 cm height grounded copper cylinder; (b) liftoff of propeller (7) at -36 kV. The last two points on the chart were recorded after liftoff. The 6-blade propeller had two opposite blades equipped each with copper tape and one pin.

**Discussion**

The present study was focused on demonstrating the viability of rotary EHD propulsion principle across more than two orders of magnitude of propeller mass. Scaling up the system does not seem to involve dramatic increase in the voltage needed for liftoff, which is also suggested by the quadratic dependence of thrust on the voltage – equation (4). Propeller up and down oscillatory motion on the HV pin can sometimes be observed before liftoff happens as also captured in the rotational speed changes - Fig. 4, (a). This is likely due to the variations in the corona current changes induced by variable air gap resistance between the propeller and the HV pin and its relation to the thrust (2). It is also in accordance to the underdamped response corresponding to equation (5). An overdamped response is generated in the case of propeller (10) Fig. 4, (b). In contrast to classic electrostatic lifters, rotary EHD devices average the EHD forces providing a smooth vertical thrust even in non-uniform distributions of the electric field (for instance near a grounded cylinder's external wall). Although the EHD thrust was not specifically measured in our experiments, a liftoff point marks the minimum vertical thrust obtained at the weight of the propeller. Using power supply readings for voltage and current, lower limits for thrust efficiency at liftoff can be obtained from relation (6). Experimental efficiency ranged for our tests from 0.64 to 6.23 N/kW in comparison to typical turbojet engines[3] at 2.5 N/kW. The highest values of 5.4 and 6.23 were obtained for the heaviest and the lightest propeller respectively. Propellers (10) and (6) have radii of 12.5 cm and 12 cm respectively. If propeller (10) could be manufactured at average mass per unit blade length of propeller (6) it would have a mass of 5.625 g. This shows within the assumption that 22.175 g could be considered as a load for propeller (10) and it gives a ratio of almost 4 between vertical thrust and equivalent propeller weight. The EHD propellers can be further optimized with respect to the number of blades, surface area, pitch, blade electrode and counter-electrode design, and dielectric, electrode and blade materials and their distribution in the system. The EHD rotary system studied is an electric motor. Scaling up the system may generate significant torque. More thrust and torque would be expected if air, selected gas[25], or gas mixtures at higher than atmospheric pressure are used[26] in encapsulated systems. Other application of EHD rotary systems could involve small EHD fast spinners calibrated as sensors for quantities on which corona current depends on.

**Methods**

**EHD propeller design.** Electrode design is essential for proper steering of the ions to produce thrust, torque, and propeller liftoff. Fig.1a,b shows the basic design used. The conductive material is aligned along the trailing edge of the blades. Conductive material can be metal tape, conductive paint, or a conductive wire. Copper foil tape with double conductive adhesive of 0.035mm thickness was often employed for propeller electrode design. Bare conductive electric paint and tungsten wire (0.0635 mm and 0.0889 mm in diameter) were also used as electrode components. Regular metal pins were cut to about 1 cm length and used either with sharp or blunt ends, both designs supporting flight. The pins were electrically connected to the copper tape or metal wire set along at least one propeller blade. 3M™ PTFE Film Electrical Tape 61 was used for partial covering of the conductive material. However, propeller liftoff was also obtained with no insulating tape. At (or towards) the end of each blade tip a metal pin about 1 cm long (connected to the conductive material on the blade) can be attached orthogonally to the local rotational radius. Frictional torque $\tau_{F_f}$ is minimized by balancing the propeller on the sharp vertical tip of a metal pin, which injects high voltage (HV) in the blades through the conductive material running along the blades, as shown in Fig. 1b. In a different setup, a metal bead is placed coaxially on the HV pin and sustains the weight of the propeller so that it can spin or slide on the vertical pin. High voltage was applied from Glassman power supplies (+60 kV - PS/FR60P05.0, -60 kV - PS/FR60N05.0).

**Counter-electrode.** In order to make an EHD propeller spin, an intense electric field must be created on the propeller electrode. This can be achieved with a large variety of counter electrodes and distances between propeller electrode and counter electrode(s). In our experiments, counter electrodes were metal cylinders, disks, and parallel plates which all supported liftoff for at least one type of propeller in Table 1. Although electrode material may play a role, similar size of copper, aluminum, and steel cylinders did not result in significant rotational speed changes. However, if the counter electrode is covered by insulating paint, the propeller may or may not spin at all.
The propeller was usually placed coaxially inside a grounded cylindrical electrode. Nevertheless, rotational motion and liftoff can be obtained even with the propeller outside the cylinder (beneath, above or sideways) although at larger voltages.

**Rotational speed measurements.** A Photron high speed camera FASTCAM SA-X2 1000K-M4 - Monochrome 1000K with 64GB memory was used for recording the rotational motion of the propeller. The frame rates ranged from 3,000 to 30,000 fps. Determination of the rotational speed (Fig. 4) was manually performed by analyzing individual video frames.

**Data availability.** The authors declare that the data supporting the findings of this study are available within the paper and its Supplementary Information file. All additional raw and derived data that support the claims within this study are available from the corresponding author upon reasonable request.

**Acknowledgements**

The authors acknowledge the contributions related to this work to the flowing undergraduate students: Joe Cesta, Justin D'Antonio, Nicholas Curinga (made the first propeller able to liftoff), Justin Ross, Thomas Jackson, Omar Attia, Alec Suites, and Edgar Solis. Also, the author is thankful for technical support to Dennis Quill, Dr. Dale Zych and to SUNY Oswego and ECE Department for partial financial support with the project.

**Author contributions**

AI has conceived most of the experiments and conducted most of them. CM conceived some experiments and contributed with technical expertise. Both AI and CM have equally contributed to the preparation of the manuscript and figures.

**Supplementary information**

The video clip demonstrating liftoff of EHD propellers includes the sequence of propeller types (4), (10), (2), (4), and (7) respectively (as labeled in Table 1).